\documentclass[12pt]{article}
\usepackage{fullpage}
\usepackage{epsf}
\usepackage{epsfig}
\usepackage{amsthm}
\usepackage{amsfonts}
\usepackage{color}
\usepackage{amsmath}
\usepackage{setspace}
\usepackage{natbib}
\usepackage{hyperref}
\usepackage{algorithmic}
\usepackage[boxed]{algorithm}
\usepackage[usenames, dvipsnames]{xcolor}

\usepackage{graphicx}
\usepackage{caption}
\usepackage{subcaption}
\usepackage[normalem]{ulem}

\begin{document}


\title{\vspace{-1cm}
A shiny update to an old experiment game}
\author{
Robert B.~Gramacy\\
Department of Statistics\\
Virginia Tech\thanks{{\em Contact:} {\tt rbg@vt.edu} or Hutcheson Hall (MCO439),
250 Drillfield Drive, Blacksburg, VA 24061}}
\date{}
\maketitle

.




\begin{abstract}
Games can be a powerful tool for learning about statistical methodology.
Effective game design involves a fine balance between caricature and realism,
to simultaneously illustrate salient concepts in a controlled setting and
serve as a testament to real-world applicability. Striking that balance is
particularly challenging in response surface and design domains, where
real-world scenarios often play out over long time scales, during which
theories are revised, model and inferential techniques are improved, and
knowledge is updated. Here I present a game, borrowing liberally from one
first played over forty years ago, that attempts to achieve that balance while
reinforcing a cascade of topics in modern nonparametric response surfaces,
sequential design and optimization. The game embeds a blackbox simulation
within a {\tt shiny} app whose interface is designed to simulate a realistic
information-availability setting, while offering a stimulating, competitive
environment wherein students can try out new methodology, and ultimately
appreciate its power and limitations. Interface, rules, timing with course
material, and evaluation are described, along with a ``case study'' involving
a cohort of students at Virginia Tech.

\end{abstract}



\section{The setting}
\label{sec:intro}

In-class games are a common way to encourage learning---to interject some fun
and build intuition in an seemingly esoteric, or tedious technical landscape.
A good game could be fundamental to retaining students, say in introductory
statistics.  One fine example uses chocolate chip cookies to illustrate
aspects of sampling distributions \citep{lee:2007}.  The long arc of an
out-of-class game played over an entire semester is attempted rather less
frequently. However for some topics, like experimental design and response
surface optimization, that setting is quite natural: real-life applications
play out on longer temporal scales, and in an inherently dynamic landscape. In
this article I present such a game, designed to simulate a real-world setting
that students might encounter on internship, which was played during a
graduate course I recently gave at Virginia Tech. The game is an update of one
first played over forty years ago \citep{mead:freeman:1973}.

\citeauthor{mead:freeman:1973}'s game was ahead of its time.  Today, few are
aware of the contribution. Although it is cited prominently in
\cite{box:draper:1987}, which is how I found it, that is one of just eight
references in the literature. Perhaps this is because, for many decades
(70s-90s, say) the setup of the game, requiring a custom computing environment
with student access, was hard to replicate.  Today, with {\sf R}/CRAN
\citep{cranR} implementation and {\tt shiny} \citep{shiny} web interfaces,
barriers have come way down.
  
The original game involves {\em blackbox} evaluation of agricultural yield
as a function of six nutrient levels, borrowed from \cite{nelder:1966} and
reproduced in {\sf R} as follows.

{\singlespacing
\begin{verbatim}
yield <- function(N, P, K, Na, Ca, Mg)
{
  l1 <- 0.015 + 0.0005*N + 0.001*P + 1/((N+5)*(P+2)) + 0.001*K + 0.1/(K+2)
  l2 <- 0.001*((2 + K + 0.5*Na)/(Ca+1)) + 0.004*((Ca+1)/(2 + K + 0.5*Na))
  l3 <- 0.02/(Mg+1)
  return(1/(l1 + l2 + l3))
}
\end{verbatim}}

\noindent Players obtain noisy yield evaluations, due to additive block and
plot-within-block (Gaussian) effects, with the ultimate goal of maximizing
yield.  There are five computer sessions (simulating crop years).  Multiple
experiments can be undertaken in a single session but strategies can only be
revised between sessions. My updated version still focuses on maximizing yield
through a series of runs, and its inaugural run involved a {\tt yield} form
very similar to the one coded above. The specifics are detailed in Section
\ref{sec:game}. But the environment no longer considers blocks and plots.

My revisions to the game are primarily motivated by modern
technology/application, and a desire to teach a more sophisticated statistical
toolkit. Classical response surface methods and design emphasize low degree
(first- and second-order) linear modeling.  The resulting steepest ascent and
ridge analyses \citep[see, e.g.,][]{box:draper:1987,myers:etal:2016} enjoy a
high degree of analytical tractability. Many relevant calculations can either
be performed by hand, or with a graphing calculator.  Yet such conveniences
offer negligible advantage in modern applications demanding facilities that,
historically, would have been considered supercomputing but are now simply
routine numerics.  Two prominent examples are information
technology/e-commerce and computer simulation experiments. 
Modern response surface methods in those domains borrow heavily from
geostatistics and machine learning with Gaussian processes, deep neural
networks, and regression and classification trees.  Sequential design
strategies like expected improvement \citep{jones:schonlau:welch:1998} promise
a more modular approach to (so-called Bayesian) optimization, allowing fancy
models to be swapped in and out, and a degree of human-free automation with
light coding.  My teaching and research philosophy favors modularity in
implementation, and the game was in part conceived to reinforce those aspects
in practice.

Several aspects of the new game setup emphasize friendly competition as a
means of enhancing the learning arena. Exploring a diverse, state-of-the-art,
toolkit benefits from sandboxing aspects of game play, but the superlative
nature of the goal---of optimizing yield---organically encourages students to
adapt and improve over time.  Relative benchmarks also enhance the sense of
realism, albeit through somewhat artificial means.\footnote{Competition
between independent agents may not be the best way to pool efforts in a
real-world setting.}  At the same time, it remains important that students
feel compelled engage with the game in a way that is advantageous
pedagogically, which for some could come at the expense of winning.  Toward
that end, Section \ref{sec:outcomes} covers player benchmarking, timing of
methodology with lecture material, and an assessment strategy designed to
encourage regular engagement and the deployment of a wealth of tools.  Some
results from a real run of the game at Virginia Tech are provided.  Section
\ref{sec:discuss} concludes with lessons learned and ideas for future
variations.  An online supplement includes a suite of supporting codes and
other materials.

\section{Game design}
\label{sec:game}

The core of game play is facilitated by an {\sf R} {\tt shiny} app, shown in
Figure \ref{f:app}, which serves as both a multi-player portal and an
interface to the back-end database of player(s) records.  An {\tt Rmarkdown}
document, {\tt yield.Rmd}, provided with the supplementary material, compiles
a full set of instructions on how to use the app, the rules of the game, and
some suggestions about strategy.  An HTML rendering of that document is
provided at \url{http://bobby.gramacy.com/teaching/rsm/yield.html}.  The
salient details follow.

\subsection{Using the {\tt shiny} app}

\begin{figure}[ht!]
\centering
\includegraphics[scale=0.31]{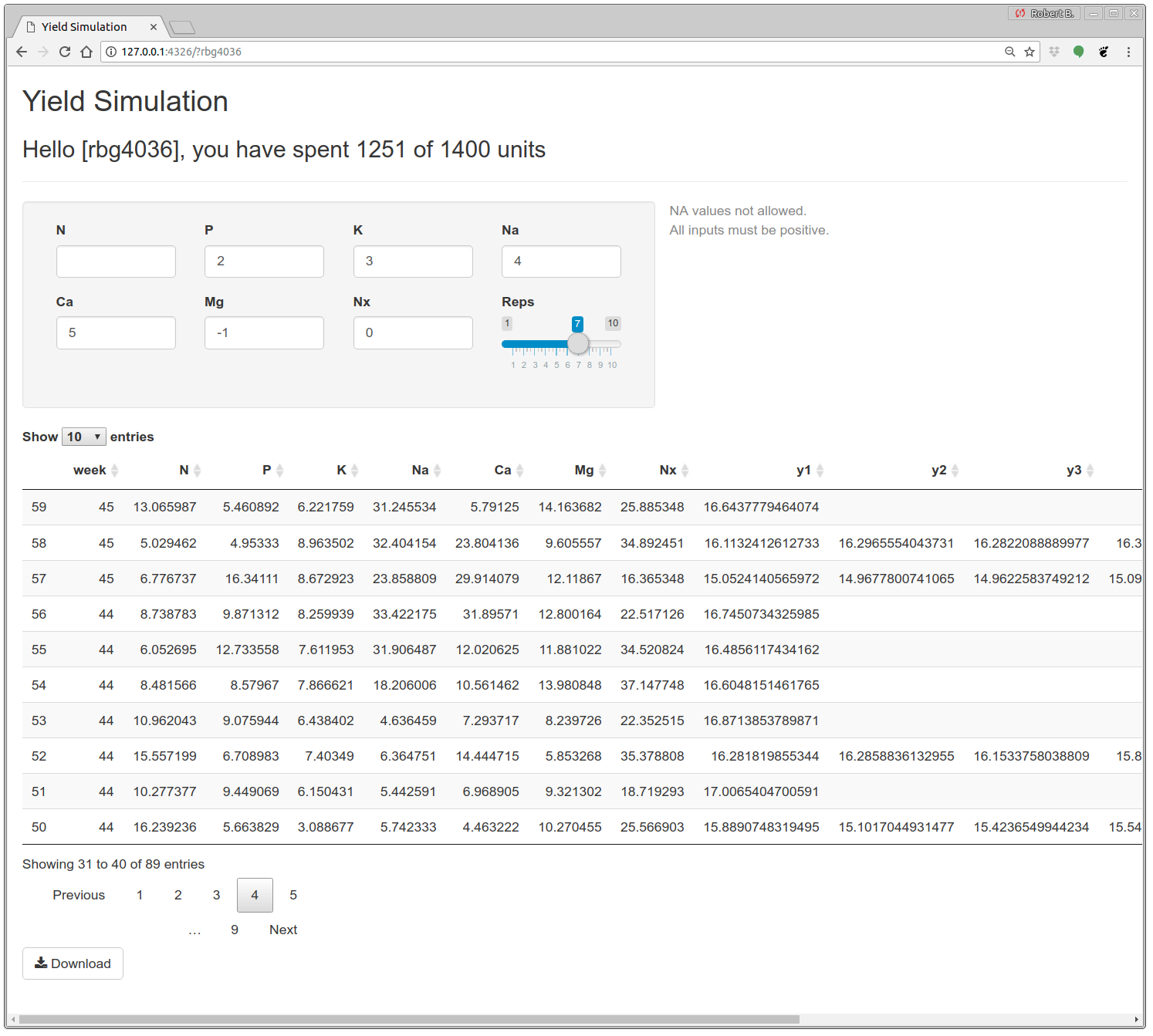}
\caption{An interactive {\tt yield} simulation session. User {\tt rbg4036} may
inspect the run budget, perform new runs (if budget remains), browse
historical runs, or download a text file.}
\label{f:app}
\end{figure}

Since all game players use the same app, interaction begins with a ``logging
in'' phase.  In advance of opening the game, I asked each player to provide me
with their initials (2--4 characters) and a four-digit secret PIN, the
combination of which comprises of the login token.  Logging in involves
providing ``url?token'' in the browser search bar.  In Figure \ref{f:app} the
URL points to a local {\tt shiny} server, and the token is ``rbg4036'', a
combination of my initials (I played the game too) and my office number, 403G.

Once logged in, the player is presented with three blocks of game content. A
greeting block provides details on spent and total budget for experimental
runs.  How the budget works, costs of runs, etc., is discussed in detail in
Section \ref{sec:rules}.  As long as the player has not fully spent, or
over-spent, their budget, new runs may be performed, via the second block on
the page. Here, the player enters the coordinates of the next run.  Until all
fields are populated with valid (positive, non-empty) values, helpful error
messages appear on the right.  The last input, which is a slider, indicates
the number of replicates desired---again discussed in more detail
shortly. Once all entries are valid, the error messages are replaced by a
``Run'' button and a warning that there are no do-overs.

Performing a run causes the table in the final block of the page to be
updated, with the week, inputs, and outputs of the new run appearing at the
top of the table.
The primary purpose of the table is to provide visual confirmation that the
run has been successfully stored in the player's database file. Buttons are
provided to aid in browsing, however this is not intended as the main
data-access vehicle.  A ``Download'' button at the bottom creates a text file
which can be saved via browser support---usually into the
\verb|~/Downloads| directory. Empty output fields are converted to {\tt NA}s
in the downloaded file.  

The appendix contains some details about the backend
of the {\tt shiny} app, including special considerations required
for hosting in the cloud, say on \verb!shinyapps.io!.

\subsection{Rules, startup and twists}
\label{sec:rules}

Players are encouraged to collaborate on strategy, and the development of
relevant mathematical calculations, but they may not share code or data.  They
start with a database containing an identical design of seven inputs,
each with five replicate responses whose noise structure is explained
momentarily. Students were introduced to the game in the fifth week (game week
zero), and could perform their first runs in the sixth week of a 15-week
semester. Including Thanksgiving and final exam weeks tallies thirteen weeks
of game play.

In each week, including week zero, students accrue 100 playing units to spend
on runs, with full roll-over from previous weeks.  The cost structure for runs
favors replicates.  Obtaining the first replicate costs ten units.  Replicates
two through four cost an additional three units each; five through seven cost
two, and eight through ten cost one each.  As long as a player's account is in
the black (i.e., positive balance), s/he can perform a new run with as many
replicates as desired, up to the limit of ten.  If performing that run causes
the balance to be zero, or negative, future runs will not be allowed (the run
box on the app disappears) until the following week, after 100 new units are
added.

Replication is important for two game ``twists'' designed to encourage players
to think about signal-to-noise trade-offs, and to nudge them to spend units
regularly, rather than save them all until the end of the semester.  The first
is that players are told that the variance of the additive noise on yield
simulations is changing weekly, following a smooth process in time, and that
they will need to provide an estimate of that variance over time for their
final report.  They are further warned that the variance may be increasing,
effectively devaluing unused units.  In fact the variance in week $w$ followed
a simple sinusoidal structure
\[
\sigma^2(w) = 0.1 + 0.05 (\cos(2\pi(w - w_s)/10)+1) \quad\quad \mbox{for starting week} \; w_s,
\]
 which peaks in the first and tenth week.  The second wrinkle is a seventh
input, {\tt Nx}, which is unrelated to the response.  Students are not told
that one of the inputs is useless, however the final writeup instructions ask
for sensitivity analyses for the inputs, including main, partial dependence,
and total effects.

\section{Timing, outcomes and evaluation}
\label{sec:outcomes}

The class was offered as a 6000-level seminar, which graduate students in
statistics usually take after they've completed the bulk of their coursework
requirements for a terminal degree (Masters or Ph.D.).  Most students were,
therefore, not novices in the core pillars of the course: linear and nonlinear
regression, design, etc.  But few had previously worked in an environment
similar one synthesized by the game. Starting in the fifth week of the
semester allowed for ramp-up on methodological training, giving students the
chance to learn/review fundamentals like steepest ascent and ridge analyses
\citep[e.g.,][Chapters 5--6]{myers:etal:2016} before entering the game as
players. Using {\sf R} code provided in class, most students were able perform
runs in the first two weeks that improved upon yields from week zero,
while getting a feel for the system. In subsequent weeks, new methodology was
introduced which students were expected to try in the game. Most students had
not, previously, been exposed to these topics, which leverage nonparametric
response surfaces.  Conceptual and implementational hurdles were anticipated
by a careful timing of training (first examples in lecture and homework) and
testing (live, in the game) stages in the course development. Details on how
game play tracks course material, cookie crumbs to ``catch up'' stragglers,
and windows into game progress are provided below.

\subsection{Subject progression and homeworks}

Homeworks were assigned roughly every two weeks and each one, from the third
onward, contained a problem on the game. Problem statements and solutions are
provided with the supplementary material. The third homework was the the most
prescriptive about what to do in the game.  It instructed the student to fit a
first-order model to the initial data set ($7
\times 5$ runs) and determine which of the seven main effects were useful for
describing variation in yield. In my own solution, only the first three were
relevant, e.g., after a backwards {\tt step}-wise selection procedure with
BIC.  Then, after reducing to a first order model having only those three
components, they were asked to search for interactions.  I found one.

With best fitted model in hand, students were asked to characterize a path of
steepest ascent, and to obtain yield simulations along that path.  This
required determining values for the remaining (in my case four) inputs. No
guidance was given here; I used a Latin hypercube sample, paired with six
settings of the three active variables a short ways along the path of steepest
ascent.  Next, students were given several options about how to proceed,
including a second-order ridge analysis, more exploration with space-filling
designs, or more steepest ascent.  In my own solution I did a bit of all
three, and the result was a second-order fit to the data that had many
relevant main effects, interactions, etc.

After the homework deadline, I released my solution so students could see what
I did, and use it to ``catch up'' with other players during subsequent weeks.
Then we transitioned to modern material involving Gaussian processes (GPs),
presented from machine learning \citep{rasmu:will:2006} and computer surrogate
modeling \citep{sant:will:notz:2003} perspectives.  Both communities
evangelize the potential for fitted GP predictive surfaces to guide searches
for global optima in blackbox functions. Machine learning researchers call
this Bayesian optimization, whereas computer modelers call it
surrogate-assisted optimization (however increasingly they are adopting the
machine learning terminology).  The simplest variation involves optimizing the
fitted GP predictive mean equations \citep{booker:etal:1999}, in lieu of
working directly with locally winnowed input--output data pairs. The method of
expected improvement \citep[EI,][]{jones:schonlau:welch:1998} and integrated
variations \citep[e.g., IECI,][]{gramacy:lee:2011} were subsequently
introduced to better balance exploration and exploitation by incorporating
degrees of predictive variability (local to global), a hallmark of statistical
decision-making. This family of nonparametric approaches were reinforced in
three questions spanning three subsequent homeworks.  Solutions were provided
(after the due dates) as plug-n-play {\sf R} scripts leveraging mature {\tt
laGP} subroutines \citep{gramacy:jss:2016}, converting a database file into a
suggested new run without any additional human interaction.

As one example consider EI, which is based on the improvement: $I(x) =
\max\{0, f^{\mathrm{min}}_n - Y(x)\}$.  The quantity $f^{\mathrm{min}}_n$ is
best value of the objective so far, and $Y(x)$ is a prediction of that
objective as a function of the inputs $x$.  Both are derived from the
estimated response surface, and are thus random variables---a property
inherited by $I(x)$. Although there are many variations in the literature, the
most common setup simplifies to treat $f^{\mathrm{min}}_n$ as a deterministic
quantity derived from minimizing mean predictive surface for $Y(x)$, and
supposes that surface to be Gaussian. In such cases, as arises when training a
GP on input-output pairs $(x_1, y_1), \dots, (x_n, y_n)$, the distribution of
$Y(x)$ is completely specified by $\mu_n(x)$ and $\sigma^2_n(x)$, which have
closed form expressions.  Integrating $I(x)$ over $Y(x)$ is also analytically
tractable in that Gaussian setting, leading to the expected improvement (EI):
\[
\mathbb{E}\{I(x)\} = (f^n_{\min} - \mu_n(x))\,\Phi\!\left(
\frac{f^n_{\min} - \mu_n(x)}{\sigma_n(x)}\right)
+ \sigma_n(x)\, \phi\!\left(
\frac{f^n_{\min} - \mu_n(x)}{\sigma_n(x)}\right),
\]
where $\Phi$ and $\phi$ are the standard normal cdf and pdf, respectively.
Notice how EI naturally balances exploitation, $\mu_n(x)$ below $f_{\min}^n$,
and exploration, large $\sigma_n(x)$, which is key to automating iterations of
search for local optima (by maximizing $\mathbb{E}\{I(x)\}$ over $x$).

Homework 5, provided in the supplementary material, asked students to try EI
in the game.  That homework was assigned in week 7 of game play, and
variations were considered in subsequent homeworks/weeks.  Although timing is
difficult to pinpoint exactly, the students' implementation of EI-like
methods, and the scripts provided as catch-up after the due date (week 9, see,
e.g., \verb!yield_gp_ei.R! in the supplementary material), are responsible for
much of the progress realized in latter weeks.  For example student ``mds'',
who had relatively mediocre success with classical RSM tools, saw big
improvements [see Figure \ref{f:leader}] from employing these more
sophisticated, but more easily automated, tools.

\subsection{Leaderboard}
\label{sec:leader}

In hopes that friendly competition would spur interest, I provided
leaderboard-style views into players' performance relative to one another,
updated in real time.  A live {\tt Rmarkdown} script compiled four views into
player progress for web viewing.
\begin{figure}[ht!]
\centering
\includegraphics[scale=0.5,trim=0 0 16 0]{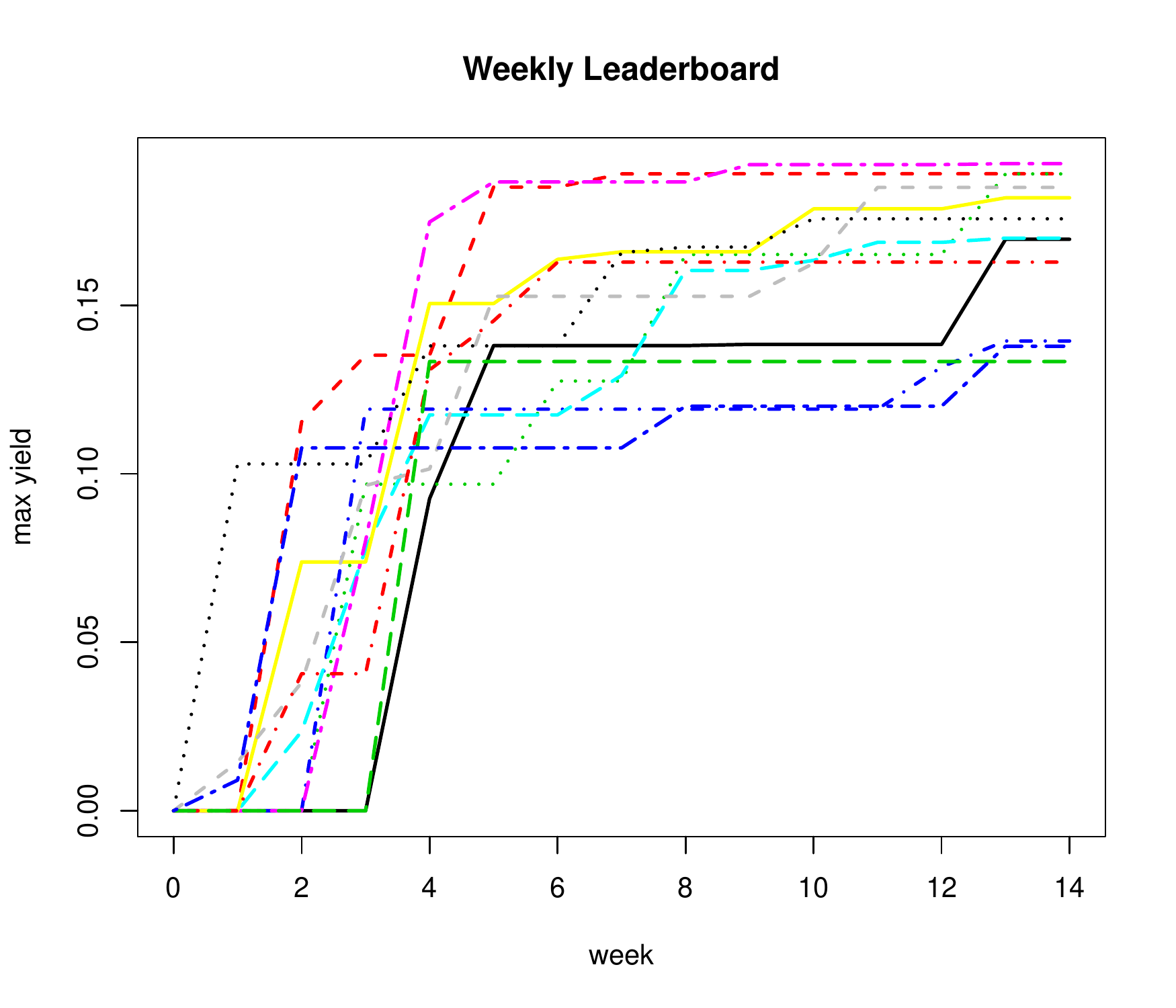}
\includegraphics[scale=0.5,trim=50 0 15 0,clip=TRUE]{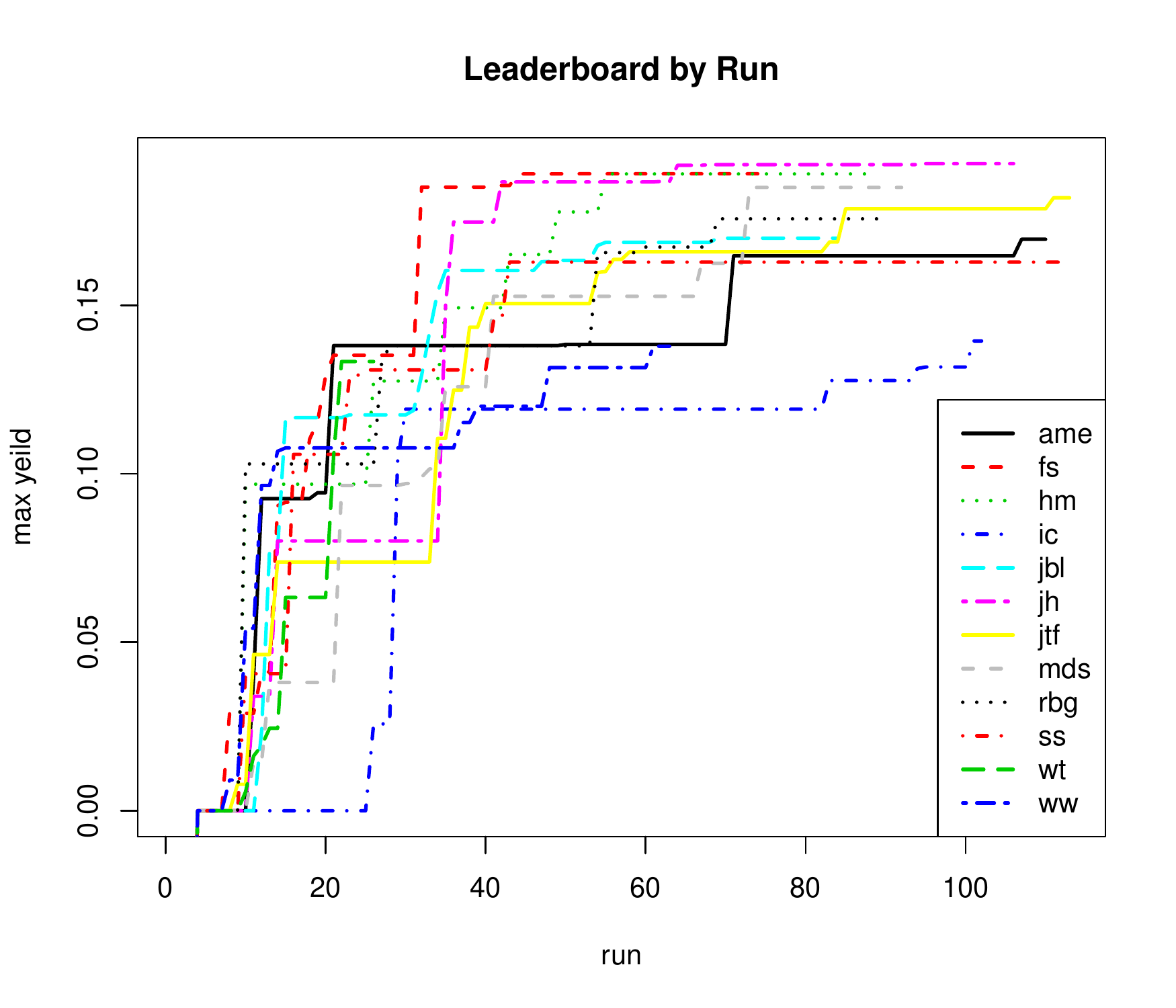}
\caption{Two de-noised views into the real-time leaderboard during the final
week of game play: best {\tt yield} by week ({\em left)}, and by run ({\em
right}).}
\label{f:leader}
\end{figure}
Two of those views, snapped during the final week of game play, are provided
in Figure \ref{f:leader}.  On the $x$-axis is the week of game play ({\em
left} panel) or the unique run number ({\em right}), and on the $y$-axis is a
normalized yield response.  Each player has a line in the plot,
with color and type indicated by their initials (masking the pin).  The
responses shown have been de-noised in order to view pure progress, making
normalization essential otherwise players would know the true mean value of
their best noisy response. The two other views provided by the {\tt Rmarkdown}
script show ``raw'' versions of these same plots, on the original
scale---these are not shown here.

Observe from the panel that about half of the progress is made in the first
five weeks, spanning around forty runs.  Students ``jh'' and ``fs'' made rapid
progress.  By contrast, ``hm'' ends up at the same place in the end, but with
more steady increments. The leaderboard would seem to partition students into
three classes, comprising of the top five, middle four, and lower four.
Evidently, student ``wt'' gave up early and ``ww'' may have misjudged the
appropriate balance of replicates versus unique runs.  These students, while
not having the strictly poorest overall performance in terms of max yield,
received the lowest grades on the project. Their final writeup reinforced an
erstwhile low engagement with many aspects of the competition, including a
failure to capitalize on the ``catch up'' scripts provided as part of the
homework keys.  Finally, observe in the figure that my own strategy placed me
fifth by these measures.  I favored replication over unique runs in hopes of
obtaining better main effects, sensitivity indices, and estimates of variance
over time.  Very few students picked up on the small asymptotic effect of {\tt
Mg}, commented on the potential useless of {\tt Nx}, or picked up on periodic
effects in the noise.  Although all three were evident in my own solution,
that's perhaps because I knew in advance what to look for.  Despite those
caveats, students with the best yield results on the leaderboard had
remarkably tight estimates for the optimal coordinates of the five ``active''
inputs, relative both to one another and to the truth.

\section{Discussion}
\label{sec:discuss}

I have described a statistical optimization game updating a previous one in
two senses.  The first sense is that the game uses modern tools for
implementation.  Supplementary materials provide {\sf R} code with support for
a {\tt shiny} app interface, cloud storage for instance-based hosting like
{\tt shinyapps.io}, and real-time views into progress via a leaderboard.
Although the original version of the game was innovative, when introduced it
was essentially unusable by others at the time owing to the nature of
computing environments available in the early 1970s.  The second update has to
do with modern response surface methods.  My re-casting of the game, and the
homework problems from class which support it, emphasize a machine learning
and computer surrogate modeling ``hands-free'' toolkit from the 21st century.

Upon reflection, the game perhaps could be revised to better emphasize these
more modern tools.  Students who made early progress on the leaderboard
unfortunately realized diminishing returns from methodological upgrades
developed later in the semester.  In some sense, the had been ``unlucky'' in
finding right answer ``too early'' to benefit from the modern tools.  I recall
an office hours session where one of the better students, ``fs'', was somewhat
distraught by lack of improvement in yield from his/her EI implementation.
Eventually, I had no choice but to reveal that s/he had essentially ``solved
it'' so that s/he could move productively on to other things. In future play I
may opt for a multi-modal blackbox in order to keep students engaged,
and to demonstrate the explorative value that comes from EI and IECI-like
heuristics appearing later in the syllabus. Another potentially exciting
variation may entertain a real blackbox simulation, and perhaps one not
involving agriculture, which could feel somewhat old-fashioned to the younger
generation of data scientists.  A promising example is the so-called
assemble-to-order \citep[ATO,][]{xie:frazier:chick:2012} solver. This
eight-dimensional example is challenging because it has a spatially dependent
noise structure.  

Future versions of the class will cover heteroskedastic GP methodology
\citep{Binois2018,Binois2018b}, via the {\tt hetGP} package for {\sf R}
\citep{Binois2017}, in order to accommodate that kind of process.  In its
inaugural run, the game featured mild heteroskedasticity in time (i.e., over
game weeks) in order to encourage regular game-play.  The number of players
performing runs in each week were 3, 10,  9, 10,  9,  9,  8, 10,  9,  9,  8,
8, and 10 (of 12 total), respectively.  Although not a controlled experiment,
these results suggest that that aspect of the game was a success. With
homeworks due (at most) by-weekly, it would be surprising to have 80\% or
higher engagement in weeks without homework due barring extra incentives.  In
the presence of spatial heteroskedastic effects, which is arguably more
realistic than sinusoids in time, an alternate incentive for regular
engagement may be needed.

\subsubsection*{Acknowledgments}

I am grateful to  students and colleagues at Virginia Tech who helped curate,
and participate in, my class on modern response surfaces and computer
experiments.  Thanks to the TAS editorial team for many thoughtful comments.

\appendix

\section{Backend details}

Under the hood, the app maintains the player database as files quite similar
to those offered for downloading.  If the app is being hosted on a standalone
node running a {\tt shiny} server, those files may be stored locally in the
app's working directory.  However, if the game is hosted in the cloud, say on
\verb!shinyapps.io!, the database can only temporarily be stored locally,
while that instance is active. Ensuring that each of multiple running
instances share the same game database,  and guaranteeing its integrity after
instances time-out due to inactivity (which causes the running directory to be
purged), requires that the files be stored elsewhere, in a single persistently
available location in the cloud. I hosted my game at
\url{http://gramacylab.shinyapps.io/yield}, with database files at
\url{www.dropbox.com} accessed through the {\tt rdrop2} \citep{rdrop2} API.
When a new instance is created, startup code triggers {\tt rdrop2} calls to
\verb!drop_get! files into the local working directory,\footnote{All of the
player files are downloaded as part of the instance initialization, which can
unfortunately be time consuming. However, the instance is then ready to serve
multiple players, if needed.} and subsequently \verb!drop_upload! to sync
those with new runs.  Note that the {\tt Rmarkdown} leaderboard would also
pull from the same {\tt rdrop2} API.  See the supplementary materials for
more details.  The default setup of those codes accomodates local (i.e.,
non-cloud) implementation, since that setup most expediently facilitates
development and other tinkering.  Commented code and description therein
clarifies how to engage the {\tt rdrop2} features for production in the cloud.

\bibliography{yield}
\bibliographystyle{jasa}

\end{document}